\documentclass[12pt,aps,showpacs]{revtex4}%
\usepackage{amsmath}
\usepackage{amssymb}
\usepackage{amsfonts}
\usepackage{graphicx}%
\providecommand{\U}[1]{\protect\rule{.1in}{.1in}}
\providecommand{\U}[1]{\protect\rule{.1in}{.1in}}
\providecommand{\U}[1]{\protect\rule{.1in}{.1in}}
\ifx\pdfoutput\relax\let\pdfoutput=\undefined\fi
\newcount\msipdfoutput
\ifx\pdfoutput\undefined\else
\ifcase\pdfoutput\else
\ifx\paperwidth\undefined\else
\ifdim\paperheight=0pt\relax\else\pdfpageheight\paperheight\fi
\ifdim\paperwidth=0pt\relax\else\pdfpagewidth\paperwidth\fi
\fi\fi\fi
\begin{document}
\title{Two Fluid Shear-Free Composites}
\author{J.P. Krisch and E.N. Glass}
\affiliation{Department of Physics, University of Michigan,\ Ann Arbor, MI }
\date{28 July 2013}

\begin{abstract}
Shear-free composite fluids are constructed from two Letelier rotated
unaligned perfect fluids. The component fluid parameters necessary to
construct a shear-free composite are investigated.\ A metric in the
Stephani-Barnes solution family and a simple stationary metric are discussed.

\end{abstract}

\maketitle

\section{Introduction}

Complex fluids with properties\ such as anisotropy and heat flow, can often be
modeled by several component perfect fluids in relative motion \cite{Let80,
HS97, AC05, LMA11}. The components can be real separate fluids with different
properties or, as Andersson and Comer \cite{AC10} have suggested, could
separately model the properties of a single fluid. Multi-fluid models are used
to describe small astrophysical systems \cite{Gho08, LAC08,HRV08,Cha08,ZP00}
as well as systems, like our Universe, that contain multiple matter content.
That content appears in combinations of radiation, particle matter, dark
energy, dark matter, and other more exotic
materials,\ \cite{GBT04,PSF05,GF06,CLP08,AFM09,HL11} and is also useful in
relativistic fluid dynamics \cite{KG11,CZ99,GN00,Cis01, ZQ04,HSW08}.\ While
three and higher \cite{Wil11,Kos11,MP11, Dun91} composite fluid models have
been suggested, the simplest anisotropic composite stress-energy is written as
a sum of two perfect fluids.\ Letelier \cite{Let80} suggested a way of
combining two perfect fluids that provides an anisotropic fluid with no heat
flow. His model is enforced by an assumed relation between the component
equations of state.\ The method will accommodate heat flow without this
assumption \cite{KG11},\ but there is an equation of state for the composite
fluid that represents a generalization of the usual pressure isotropy
condition\ \cite{Ber81,SR84,Den89}. How the fluid parameters of the composite
are driven by the parameters of the component fluids is important, given the
recent interest in shear-free perfect fluids \cite{Ell11} and the shear-free
fluid conjecture \cite{Col86,Sop98,VdB99,VCK07,CKV+09}.\ With the increasing
use of multi-fluid models, understanding the effects caused by component fluid
motions can provide insights to the origin of composite fluid velocity
components such as shear, expansion, and vorticity.

In this paper we construct shear-free composites from perfect fluids and
examine how the composite fluid stress-energy and metric depend on the
component fluid alignment. The component fluids are combined using the
Letelier method with an alignment condition and this is briefly reviewed in
the next section.\ The component/composite fluid parameters and stress-energy
are connected to the metric for the containing spacetime. In the third
section, we consider two metric examples. The first metric is taken from the
Stephani-Barnes solution family \cite{Kra97} and uses two shear-free
components. The second example describes a simple rotating spacetime and its
concomitant vorticity.\ Three Appendices relating the composite/component
fluid \ properties are provided.

\section{Two Fluid Formalism with Alignment}

\subsection*{Transformations and Composite Stress-Energy}

Consider a manifold which contains two perfect fluids and metric $g_{ab}$. A
composite stress-energy for the perfect fluids is\
\begin{equation}
T_{ab}=(\varepsilon_{1}+p_{1})U_{a}^{(1)}U_{b}^{(1)}+(\varepsilon_{2}%
+p_{2})U_{a}^{(2)}U_{b}^{(2)}+(p_{1}+p_{2})g_{ab}. \label{Stress-en1}%
\end{equation}
A single non-perfect fluid stress-energy tensor can be constructed by
transforming the two timelike component fluid velocities $[U_{i}^{(1)}%
,U_{i}^{(2)}]$ into an unnormed timelike, spacelike pair\ $[U_{i}^{\ast
},\Upsilon_{i}^{\ast}]\ $related to the unit pair, $[\hat{U}^{i},\hat
{\Upsilon}^{i}]$ by $N_{u}\hat{U}^{i}=U^{i\ast}$ and $N_{\Upsilon}%
\hat{\Upsilon}^{i}=$\ $\Upsilon^{i\ast}$.\ The composite fluid will be
described by a tetrad $[U^{a},\Upsilon^{a},X^{a},Z^{a}],$ where $X^{a}$ and
$Z^{a}$ are any two spatial vectors orthogonal to $\Upsilon^{a}$. The basic
velocity transformations with inverse are%
\begin{equation}%
\begin{bmatrix}
N_{u}\hat{U}_{a}\\
N_{\Upsilon}\hat{\Upsilon}_{a}%
\end{bmatrix}
=%
\begin{bmatrix}
\cos\alpha & B\sin\alpha\\
-B^{-1}\sin\alpha & \cos\alpha
\end{bmatrix}%
\begin{bmatrix}
U_{a}^{(1)}\\
U_{a}^{(2)}%
\end{bmatrix}
\label{norm-vecs}%
\end{equation}%
\begin{equation}
\text{ \ }\ \
\begin{bmatrix}
U_{a}^{(1)}\\
U_{a}^{(2)}%
\end{bmatrix}
=%
\begin{bmatrix}
\cos\alpha & -B\sin\alpha\\
B^{-1}\sin\alpha & \cos\alpha
\end{bmatrix}%
\begin{bmatrix}
N_{u}\hat{U}_{a}\\
N_{\Upsilon}\hat{\Upsilon}_{a}%
\end{bmatrix}
\label{U-vecs}%
\end{equation}
The normalization factors are
\begin{align}
N_{\Upsilon}^{2}  &  =\Upsilon^{a\ast}\Upsilon_{a}^{\ast}=B^{-2}(\sin
^{2}\alpha-B^{2}\cos^{2}\alpha)/\cos2\alpha,\\
N_{u}^{2}  &  =-U^{\ast a}U_{a}^{\ast}=(\cos^{2}\alpha-B^{2}\sin^{2}%
\alpha)/\cos2\alpha.\nonumber
\end{align}
The general rotation angle, $\alpha,$ is fixed with terms of the component
velocity overlap by requiring $\Upsilon^{a}U_{a}=0,\Upsilon^{a\text{ }}%
$spacelike and $U^{a}$ timelike.%
\begin{equation}
\digamma=U^{(1)a}U_{a}^{(2)}=-\frac{1-B^{2}}{2B}\tan2\alpha\label{B-condition}%
\end{equation}
The overlap, $U^{(1)a}U_{a}^{(2)}$, between the component four-velocities is
assumed constant. Substituting $U_{a}^{(1)}$and $U_{a}^{(2)}$ from
Eq.(\ref{U-vecs}) into stress-energy Eq.(\ref{Stress-en1}), the composite
stress-energy is%
\begin{equation}
T_{ab}=(\varepsilon+\Pi)\hat{U}_{a}\hat{U}_{b}+(P-\Pi)\hat{\Upsilon}_{a}%
\hat{\Upsilon}_{b}+\Pi g_{ab\ }+Q_{a}\hat{U}_{b}+Q_{b}\hat{U}_{a}
\label{Stress-en2}%
\end{equation}
with components
\begin{subequations}
\label{stress-en2-comps}%
\begin{align}
\Pi &  =p_{1}+p_{2}\\
\varepsilon+\Pi &  =N_{u}^{2}[(\varepsilon_{1}+p_{1})\cos^{2}\alpha
+B^{-2}(\varepsilon_{2}+p_{2})\sin^{2}\alpha]\\
P-\Pi &  =N_{\Upsilon}^{2}[B^{2}(\varepsilon_{1}+p_{1})\sin^{2}\alpha
+(\varepsilon_{2}+p_{2})\cos^{2}\alpha]\\
Q_{a}  &  =Q\hat{\Upsilon}_{a},\text{ \ \ }Q=N_{u}N_{\Upsilon}[B^{-1}%
(\varepsilon_{2}+p_{2})-B(\varepsilon_{1}+p_{1})]\sin\alpha\cos\alpha
\end{align}
The anisotropic stress is $P$, and $Q_{a}$ is the heat current associated with
$\hat{\Upsilon}_{a}$.\ Letelier \cite{Let80} defined $B=\sqrt{\frac
{\varepsilon_{2}+p_{2}}{\varepsilon_{1}+p_{1}}}$\ in order to zero the off
diagonal stress-energy heat flow term, leaving a 1-parameter description of
the composite fluid. A convenient way to define $B$ that maintains the heat
flow is $B=\tan\alpha_{0}$ \cite{KG11}. With this choice of $B$, the velocity
overlap equation, Eq.(\ref{B-condition}) can be rewritten.
\end{subequations}
\begin{equation}
\digamma=U^{(1)a}U_{a}^{(2)}=-\frac{\tan2\alpha}{\tan2\alpha_{0}}%
\end{equation}
The fluids are aligned when $\alpha=\alpha_{0}$. To maintain the
spacelike/timelike nature of $\Upsilon^{a\ast}$ and $U^{a\ast},$ the range for
$(\alpha,\alpha_{0})$ is $0<(\alpha,\alpha_{0})<\pi/4$.\ For $\alpha
=\alpha_{0},$ with two aligned components, the equation of state simply gives
the isotropic stress, $P=\Pi.$ For aligned fluids $N_{\Upsilon}=0$; the heat
flow vanishes and the composite is another isotropic perfect fluid. \ 

\subsection*{Composite Equation of State}

The alignment restriction of $B=\tan\alpha_{0}$ removes the equation of state
assumption on the component fluids, allows non-zero heat flow, and links the
fluid description to the alignment of the component velocities.\ The
definition of $B=\tan\alpha_{0}$ does impose an equation of state on the
system. Unlike the Letelier assumption, the imposed equation of state involves
the composite fluid parameters and follows from the definitions of those
parameters.%
\begin{equation}
(\varepsilon+\Pi)S^{2}(\alpha,\alpha_{0})+(\Pi-P)C^{2}(\alpha,\alpha
_{0})+2Q\cot(2\alpha)S(\alpha,\alpha_{0})C(\alpha,\alpha_{0})=0
\label{Let-eos}%
\end{equation}
Eq.(\ref{Let-eos}) is an identity in term of the component parameters and
imposes no component conditions. It can be used as a differential equation for
the metric functions or replace the field equation for the heat scalar. \ The
two functions, $C^{2}(\alpha,\alpha_{0})$ and $S^{2}(\alpha,\alpha_{0})$ are
related to the normalization constants%
\begin{subequations}
\begin{align}
C^{2}(\alpha,\alpha_{0})  &  =N_{u}^{2}\cos(2\alpha)\cos^{2}(\alpha_{0})\\
S^{2}(\alpha,\alpha_{0})  &  =N_{\Upsilon}^{2}\cos(2\alpha)\sin^{2}(\alpha
_{0})
\end{align}
For $\alpha=\alpha_{0}$, the equation of state gives the isotropic stress
result,\ $P=\Pi.$ The size of the stress anisotropy rises as the fluids become
more and more unaligned. When the field equations are substituted for the
fluid parameters, this equation of state provides a generalization of the
metric condition imposed by stress isotropy \cite{Ber81,SR84,Den89}.\ The
stress-energy following from the field equations is implicit in the equation
of state,\ but it can be useful in providing an alignment related differential
equation for the metric potentials, or used to replace a field equation.\ In
the next section we give two examples. \ 

\section{Metric Examples}

\subsection*{Stephani-Barnes metric}

\ In this section, we consider the spherically symmetric metric in the
Stephani-Barnes \cite{Str68,Kra97} family and examine the properties of the
composite fluid that could result from two shear-free perfect fluid components
with a velocity in the radial direction. \ \ \
\end{subequations}
\begin{equation}
ds^{2}=L(t,r)^{-2}[-N(t,r)^{2}dt^{2}+dr^{2}+r^{2}d\Omega^{2}] \label{S-B-met}%
\end{equation}
The single composite fluid has a co-moving four velocity
\begin{equation}
U^{a}=[\frac{L(t,r)}{N(t,r)},0,0,0]
\end{equation}
The composite velocity is shear-free with expansion $\theta=-3\dot{L}%
/N.$\ From Eq.(\ref{U-vecs}), the component four velocities that can build the
composite are written as%
\begin{equation}
U_{\ }^{(i)a}=A^{(i)}\hat{U}^{a}+C^{(i)}\hat{\Upsilon}^{a}%
\end{equation}
with
\begin{align}
A^{(1)}  &  =\cos\alpha N_{u},\text{ \ \ \ \ }C^{(1)}=-B\sin\alpha
N_{\Upsilon}\\
A^{(2)}  &  =B^{-1}\sin\alpha N_{u},\text{ \ }C^{(2)}=\cos\alpha N_{\Upsilon
}\nonumber
\end{align}
While the composite is shear-free, the component velocities, with spatial
components, in general have shear.\ Choosing $\Upsilon^{a}$ in the radial
direction, the component fluid velocities are
\[
U_{a}^{(i)}=[-\frac{A^{(i)}N}{L},\frac{C^{(i)}}{L},0,0].
\]
\ For this velocity the shear tensor components of the component fluids all
depend on the same term
\begin{equation}
\sigma_{ab}\sim(\frac{N,_{r}}{N}-\frac{1}{r})
\end{equation}
For the components to be shear-free, $N(t,r)$ must be separable, with form
\begin{equation}
N(t,r)=rN(t)
\end{equation}
Absorbing the factor $N(t)$ into the time coordinate, the metric that could
contain shear-free components with radial spatial velocities is%
\begin{equation}
ds^{2}=L(t,r)^{-2}[-r^{2}dt^{2}+dr^{2}+r^{2}d\Omega^{2}]. \label{L-met}%
\end{equation}
This metric is a special case of a metric considered by Sussman and reviewed
by Krasinski \cite{Sus93,Kra97}. The metric form in Eq.(\ref{L-met}) has been
discussed in the literature under a number of special conditions
\cite{Str68,Ber81,SR84,Den89}.

\subsubsection{Stress Energy}

The field equations provide the composite stress-energy components
\begin{subequations}
\label{stress-en3-comps}%
\begin{align}
8\pi\varepsilon &  =\frac{3(L_{,t})^{2}}{r^{2}}+L^{2}[2\frac{L,_{rr}}%
{L}-3(\frac{L,_{r}}{L})^{2}+\frac{4}{r}(\frac{L,_{r}}{L})]\\
8\pi\Pi &  =\frac{L^{2}}{r^{2}}[2\frac{L,_{tt}}{L}-3(\frac{L,_{t}}{L}%
)^{2}]+L^{2}[-2\frac{L,_{rr}}{L}+3(\frac{L,_{r}}{L})^{2}-\frac{4}{r}%
(\frac{L,_{r}}{L})+\frac{1}{r^{2}}]\\
8\pi P_{r}  &  =\frac{L^{2}}{r^{2}}[2\frac{L,_{tt}}{L}-3(\frac{L,_{t}}{L}%
)^{2}]+L^{2}[3(\frac{L,_{r}}{L})^{2}-\frac{6}{r}(\frac{L,_{r}}{L})+\frac
{2}{r^{2}}]\\
Q_{r}  &  =-2(\frac{L,_{t}}{r})_{,r}%
\end{align}
Substituting the fluid parameters into the equation of state from the Letelier
composition Eq.(\ref{Let-eos}), one finds%
\end{subequations}
\begin{equation}
\frac{2L,_{tt}}{L}-\frac{4r^{2}}{L}\partial_{r}(\frac{L,_{t}}{r})\cot
(2\alpha)\frac{C(\alpha,\alpha_{0})}{S(\alpha,\alpha_{0})}+2r^{2}(\frac
{L,_{r}}{rL}-\frac{L,_{rr}}{L}-\frac{1}{2r^{2}})\frac{C^{2}(\alpha,\alpha
_{0})}{S^{2}(\alpha,\alpha_{0})}+1=0. \label{L-dt-dr}%
\end{equation}
If, for example, $L$ $=L(r)$ the equation of state is
\begin{equation}
\frac{L_{,rr}}{L}-\frac{L_{,r}}{rL}+\frac{1}{2r^{2}}=0. \label{isotropy-cond}%
\end{equation}
an Euler equation, homogeneous in $r$ of degree $-2$. The stresses are
isotropic. The additional structure in equation (\ref{L-dt-dr}) extends the
single fluid isotropy condition. The case where $L$ is only a function of
time, $L=L(t)$, is not isotropic$.$ For this case, the equation of state
becomes a simple differential equation for $L(t)$. \
\begin{equation}
\frac{2L,_{tt}}{L}+\frac{4L,_{t}}{L}\cot(2\alpha)\frac{C(\alpha,\alpha_{0}%
)}{S(\alpha,\alpha_{0})}-\frac{C^{2}(\alpha,\alpha_{0})}{S^{2}(\alpha
,\alpha_{0})}+1=0.
\end{equation}
For unaligned fluids the metric function $L(t)$ is an exponential sum
depending on the alignment of the two perfect fluids.\ A simple exponential
solution is \
\begin{equation}
L(t)=L_{1}e^{\beta t}.
\end{equation}
$\beta$ is related to\ $(\alpha,\alpha_{0})$ through the equation of state and
satisfies the quadratic equation%
\begin{equation}
\beta^{2}+2\beta\left[  \cot(2\alpha)\frac{C(\alpha,\alpha_{0})}%
{S(\alpha,\alpha_{0})}\right]  +\left[  \frac{S^{2}(\alpha,\alpha_{0}%
)-C^{2}(\alpha,\alpha_{0})}{2S^{2}(\alpha,\alpha_{0})}\right]  =0
\end{equation}
The fluid content and curvature scalar are
\begin{subequations}
\label{Lt-results}%
\begin{align}
8\pi\varepsilon &  =\frac{L^{2}}{r^{2}}(3\beta^{2})\\
8\pi\Pi &  =\frac{L^{2}}{r^{2}}(1-\beta^{2})\\
8\pi P_{r}  &  =\frac{L^{2}}{r^{2}}(2-\beta^{2})\\
Q  &  =\frac{L^{2}}{r^{2}}(2\beta)\\
R  &  =\frac{L^{2}}{r^{2}}2(3\beta^{2}-2)
\end{align}
Positive stress requires $0<$\ $\beta^{2}<1$. This solution does not have an
isotropic limit and exhibits a curvature dependence on alignment.

\subsubsection{The component fluids}

The shear-free velocity components for this example are
\end{subequations}
\[
U_{a}^{(i)}=[-\frac{A^{(i)}r}{L},\frac{C^{(i)}}{L},0,0].
\]
with component expansions
\[
\theta_{(i)}=3\frac{-A_{(i)}L,_{t}+C_{(i)}L}{r}\
\]
In Appendix A, we examine the conditions necessary for two shear-free
components to combine in a shear-free composite. One of the conditions relates
the composite and component expansions
\begin{equation}
N_{u}\theta=-3N_{u}L,_{t}/r=\cos\alpha\theta_{1}+B\sin\alpha\ \theta_{2}%
\end{equation}

This is satisfied for these shear-free components, with the $C^{(i)}$
component contributions cancelling. An interesting insight is that the
component fluids can be expanding or contracting due to their $C^{(i)}$
contribution, while the sign of the composite expansion is determined by the
metric. A second condition is that the two component fluids must have unequal
accelerations along the direction of anisotropy. For the components in this
example, the accelerations are%
\begin{align*}
\dot{U}_{a}^{(i)}  &  =[C^{(i)}\frac{L,_{t}}{L}-A^{(i)}][C^{(i)}%
,-\frac{A^{(i)}}{r},0,0]\\
C^{(1)}\frac{L,_{t}}{L}-A^{(1)}  &  =BN_{\Upsilon}(-\sin\alpha\frac{L,_{t}}%
{L}-\frac{N_{u}}{BN_{\Upsilon}}\cos\alpha)\\
C^{(2)}\frac{L,_{t}}{L}-A^{(2)}  &  =\ N_{\Upsilon}(\cos\alpha\frac{L,_{t}}%
{L}-\frac{N_{u}}{BN_{\Upsilon}}\sin\alpha)\
\end{align*}
with clearly unequal acceleration. One of the components could be
unaccelerated. For example, if fluid 2 is unaccelerated%

\[
\frac{L,_{t}}{L}=\frac{A^{(2)}}{C^{(2)}}=\frac{N_{u}}{BN_{\Upsilon}}\tan\alpha
\]
then the second fluid also has zero expansion and the metric is determined by
the alignment conditions.

\subsection*{A Simple Rotating Metric}

The second metric example is a simple cylindrical stationary spacetime with
metric
\begin{equation}
ds^{2}=-dt^{2}-2f(r)d\phi dt+r^{2}d\phi^{2}+dr^{2}+dz^{2} \label{rot-cyl-met}%
\end{equation}
The composite fluid will be constructed from two perfect fluids with shear and
vorticity. \ \ The motion and stress energy will be decomposed using a
comoving tetrad.
\[
U_{a}=-[1,0,f,0],\text{ \ }R_{a}=[0,1,0,0],\text{ \ }\Phi_{a}=[0,0,D,0],\text{
\ }Z_{a}=[0,0,0,1].
\]
\ The stress-energy of the composite fluid is, with $D^{2}=r^{2}+f^{2}$\ \
\begin{subequations}
\begin{align}
8\pi\varepsilon &  =-\frac{D,_{rr}}{D}+3\frac{f,_{r}^{2}}{D^{2}}\\
8\pi P_{\phi}  &  =\frac{f,_{r}^{2}}{D^{2}}\\
8\pi P_{r}  &  =\frac{f,_{r}^{2}}{D^{2}}\\
8\pi P_{z}  &  =\frac{D,_{rr}}{D}-\frac{f,_{r}^{2}}{D^{2}}%
\end{align}
Since the alignment formalism describes a fluid with a single anisotropic
stress, the direction of anisotropy is $Z^{i}$ and there will be heat flow
down the axis of the cylindrical system. \
\end{subequations}
\begin{equation}
\ Q\cot(2\alpha)S(\alpha,\alpha_{0})C(\alpha,\alpha_{0})=\frac{D,_{rr}}%
{2D}[S^{2}(\alpha,\alpha_{0})+C^{2}(\alpha,\alpha_{0})]-\frac{f,_{r}^{2}%
}{D^{2}}[2S^{2}(\alpha,\alpha_{0})+C^{2}(\alpha,\alpha_{0})]\nonumber
\end{equation}

The velocity of the composite is shear and expansion free, with vorticity
$\omega_{r\phi}=f_{,r}/2$.\ The component velocities have a component in the
z-direction. \
\begin{equation}
U^{(i)a}=A^{(i)}\hat{U}^{a}+C^{(i)}Z^{a}.
\end{equation}
and can have shear and vorticity.The standard vorticity definition based on
the $U_{a}^{(i)}$ covariant derivative expansion is%
\begin{align*}
\omega_{ab}^{(i)}  &  =U_{[a;b]}^{(i)}+\dot{U}_{[a}^{(i)}U_{b]}^{(i)}\ \\
U^{(i)a}\omega_{ab}^{(i)}  &  =0
\end{align*}
The component vorticities could have a time like component with the velocity
constraint providing the relation
\[
\lbrack A^{(i)}-\frac{C^{(i)}f}{D}]\omega_{tb}^{(i)}=-\frac{C^{(i)}}{D}%
\omega_{\phi b}^{(i)}%
\]
If the fluids are aligned, $C^{(i)}=0$ $(N_{\Upsilon}=0)$, there is only the
single composite vorticity, $\omega_{r\phi},$as the components rotate with the
composite value. \ 

This description could be compared to the two-fluid model based on momentum
\cite{SLA+10,CPA11}.\ In these models the fluid current, $n^{(i)a}$, and
momentum $\mu^{(i)a},$ are defined in terms of the component velocities and
currents
\begin{align}
n^{(1)a}  &  =n^{(1)}U^{(1)a}\\
\mu_{a}^{_{(1)}}  &  =B_{(1)}n^{(1)}U_{a}^{(1)}+A_{(12)}n^{(2)}U_{a}%
^{(2)}\nonumber
\end{align}
with $B_{(1)}$ and $A_{(12)}$ constants of the formalism. The vorticity is
based on the momentum and, for $i=1$ (fluid x in their notation) is defined
as
\begin{align}
\bar{\omega}_{ab}^{(1)}  &  =\mu_{a;b}^{_{(1)}}-\mu_{b;a}^{_{(1)}}\\
U^{(1)a}\bar{\omega}_{ab}^{(1)}  &  =0
\end{align}
For the simple stationary metric of this example with only r-dependence, the
momentum based vorticities are%
\begin{subequations}
\begin{align}
\bar{\omega}_{tb}^{(1)}  &  =\partial_{b}\mu_{t}^{(1)}\\
\bar{\omega}_{\phi b}^{(1)}  &  =\partial_{b}\mu_{\phi}^{(1)}\\
U^{(1)t}\partial_{b}\mu_{t}^{(1)}  &  =-U^{(1)\phi}\partial_{b}\mu_{\phi
}^{(1)}%
\end{align}
If the velocities are aligned and taken comoving, conditions are placed on the
functional structure of the component momentum. \ The use of a velocity
constraint on a momentum based vorticity also raises questions about fluid
motion in more complicated metrics.\ Spin fluid calculations have demonstrated
different physical descriptions arising from imposing a momentum versus
velocity constraint \cite{RB95,FCN+11,FCH+12} on a spin tensor.\ 

\section{Discussion}

In summary, we have considered a shear-free anisotropic composite fluid
constructed from two perfect fluids and shown that, when the component fluid
alignment is considered, a composite equation of state is imposed which
extends the usual isotropy condition.\ We considered two examples. The first
used \ two shear-free component fluids with a radial velocity component. The
metric for the fluid provided a specific solution to a family of conformal
Killing vector metrics discussed by Sussman \cite{Sus93}. The second example
considered a simple rotating fluid.\ This example illustrated a possible
difference in the alignment description used in this paper and the
momentum/current description.

We have focused on the composite fluid produced by combining two perfect
fluids.\ However, the inverse Letelier expansion of two perfect fluids, in
terms of the composite four velocity and anisotropy vector, can be related to
a potential representation.\ For example, Schutz \cite{Sch70}, expanding the
minimal set of Clebsch potentials, introduced a six-potential representation
for a perfect fluid four velocity.\ \
\end{subequations}
\[
U_{a}=\mu^{-1}(\varphi,_{a}+\alpha_{c}\beta_{c},_{a}+Cs_{,a})
\]
$\mu^{-1}$ is the specific inertial mass.\ $s$ is the specific entropy. $C$
(usually called $\theta),$ is the thermasy of Van Danzig \cite{DvD39},
$\alpha_{c}$ and $\beta_{c}$ are the Clebsch potentials \cite{Sch70} needed
for a fluid with vorticity.\ In the Letelier expansion,\ each of the component
velocities for the anisotropic fluid is expanded with a minimal vector set
related to the component fluid alignment. \
\begin{equation}
U_{a\ }^{(i)}=A^{(i)}\hat{U}_{a}+C^{(i)}\hat{\Upsilon}_{a}%
\end{equation}
Combining the Schutz potentials with a multifluid combination method could
relate the alignment parameters to velocity potentials and provide a
broader\ multifluid formalism.

\appendix{}

\section{Relating the component fluid accelerations}

The component velocity overlap equation enforces the orthogonality of the
transformed $[U_{a},\Upsilon_{a}]$ pair.%
\[
\digamma:=U^{(1)a}U_{a}^{(2)}=-\frac{(1-B^{2})}{2B}\tan2\alpha\
\]
$\alpha=\alpha_{0}$ is the condition for $U_{a}^{(1)}$ and $U_{a}^{(2)}$ to be
aligned. $\alpha$ varies the size of the component spatial speeds.\ Assuming
constant angles the covariant derivative is%
\begin{equation}
U_{a;b}^{(1)}U^{(2)a}=-U^{(1)a}U_{a;b}^{(2)}%
\end{equation}
Expanding the derivatives and forming the products with $U^{(2)b}$ and,
separately with $U^{(1)b}$, we have%
\begin{align}
\lbrack-\dot{U}_{a}^{(1)}U_{b}^{(1)}+\omega_{ab}^{(1)}+\sigma_{ab}^{(1)}%
+\frac{\theta_{(1)}}{3}h_{ab}^{(1)})]U^{(2)a} &  =-[-\dot{U}_{a}^{(2)}%
U_{b}^{(2)}+\omega_{ab}^{(2)}+\sigma_{ab}^{(2)}+\frac{\theta_{(2)}}{3}%
h_{ab}^{(2)}]U^{(1)a}\\
\dot{U}_{a}^{(1)}U^{(2)a} &  =\dot{U}_{a}^{(2)}U^{(1)a}\digamma-\sigma
_{ab}^{(2)}U^{(1)b}U^{(1)a}-\frac{\theta_{(2)}}{3}h_{ab}^{(2)}U^{(1)b}%
U^{(1)a}\nonumber\\
\dot{U}_{a}^{(2)}U^{(1)a} &  =\dot{U}_{a}^{(1)}U^{(2)a}\digamma-\sigma
_{ab}^{(1)}U^{(2)a}U^{(2)b}-\frac{\theta_{(1)}}{3}h_{ab}^{(1)}U^{(2)b}%
U^{(2)a}\nonumber
\end{align}
Substituting we find the two equations for the acceleration
\begin{align}
\dot{U}_{a}^{(1)}U^{(2)a}(1-\digamma^{2}) &  =(\digamma\frac{\theta_{(1)}}%
{3}+\frac{\theta_{(2)}}{3})(1-\digamma^{2})-\digamma\sigma_{ab}^{(1)}%
U^{(2)a}U^{(2)b}-\sigma_{ab}^{(2)}U^{(1)b}U^{(1)a}\label{acc-vel-1}\\
\dot{U}_{a}^{(2)}U^{(1)a}(1-\digamma^{2}) &  =(\frac{\theta_{(1)}}{3}%
+\digamma\frac{\theta_{(2)}}{3})(1-\digamma^{2})-\sigma_{ab}^{(1)}%
U^{(2)a}U^{(2)b}-\digamma\sigma_{ab}^{(2)}U^{(1)b}U^{(1)a}\label{acc-vel-2}%
\end{align}
For shear-free components the expression simplifies to
\begin{align}
\dot{U}_{a}^{(1)}U^{(2)a} &  =(\digamma\frac{\theta_{(1)}}{3}+\frac
{\theta_{(2)}}{3})\\
\dot{U}_{a}^{(2)}U^{(1)a} &  =(\frac{\theta_{(1)}}{3}+\digamma\frac
{\theta_{(2)}}{3})
\end{align}

\section{Composite fluid parameters}

In this Appendix we use the Letelier transformation equations to develop
expressions for the acceleration, expansion, shear and vorticity of the
composite fluid. The transformation equations are
\begin{subequations}
\label{Beqn}%
\begin{align}
N_{u}U_{a} &  =(\cos\alpha)U_{a}^{(1)}+(B\sin\alpha)U_{a}^{(2)}\label{Ba}\\
N_{\Upsilon}\Upsilon_{a} &  =-(B^{-1}\sin\alpha)U_{a}^{(1)}+(\cos\alpha
)U_{a}^{(2)}\label{Bb}\\
U_{a}^{(1)} &  =(\cos\alpha N_{U})U_{a}-(B\sin\alpha N_{\Upsilon})\Upsilon
_{a}\label{Bc}\\
U_{a}^{(2)} &  =(B^{-1}\sin\alpha N_{U})U_{a}+(\cos\alpha N_{\Upsilon
})\Upsilon_{a}\label{Bd}%
\end{align}
Taking the covariant derivative of $U_{a}$ and expanding one obtains
\end{subequations}
\begin{align}
N_{u}(-\dot{U}_{a}U_{b}+\omega_{ab}+\sigma_{ab}+\frac{\theta}{3}h_{ab}) &
=\cos\alpha(-\dot{U}_{a}^{(1)}U_{b}^{(1)}+\omega_{ab}^{(1)}+\sigma_{ab}%
^{(1)}+\frac{\theta_{(1)}}{3}h_{ab}^{(1)})\label{B2}\\
&  +B\sin\alpha(-\dot{U}_{a}^{(2)}U_{b}^{(2)}+\omega_{ab}^{(2)}+\sigma
_{ab}^{(2)}+\frac{\theta_{(2)}}{3}h_{ab}^{(2)})\nonumber
\end{align}

\subsection*{Composite Expansion and Acceleration}

\subsubsection*{Expansion}

Taking the trace of Eq.(\ref{B2}), the composite and component expansions are
related by%
\begin{equation}
N_{u}\theta=\cos\alpha\theta_{(1)}+B\sin\alpha\ \theta_{(2)} \label{B3}%
\end{equation}

\subsubsection*{Acceleration}

The acceleration follows from the covariant derivative and we have%
\begin{align}
N_{u}\dot{U}_{a}  &  =\cos\alpha(-\dot{U}_{a}^{(1)}U_{b}^{(1)}U^{b}%
+\sigma_{ab}^{(1)}U^{b}+\omega_{ab}^{(1)}U^{b}+\frac{\theta_{(1)}}{3}%
U^{b}h_{ab}^{(1)})\\
&  +B\sin\alpha(-\dot{U}_{a}^{(2)}U_{b}^{(2)}U^{b}+\sigma_{ab}^{(2)}%
U^{b}+\omega_{ab}^{(2)}U^{b}+\frac{\theta_{(2)}}{3}U^{b}h_{ab}^{(2)})\nonumber
\end{align}
Using the transformation equations to generate the velocity dot products, the
acceleration can be rewritten
\begin{align}
\dot{U}_{a}  &  =\cos^{2}\alpha\ \dot{U}_{a}^{(1)}+\frac{B\sin\alpha\cos
\alpha}{N_{u}^{2}}[\sigma_{ab}^{(1)}U^{(2)b}+\omega_{ab}^{(1)}U^{(2)b}%
+\frac{\theta_{(1)}}{3}(U_{a}^{(2)\ }+\digamma U_{a}^{(1)})]\\
&  +\sin^{2}\alpha\ \dot{U}_{a}^{(2)}+\frac{B\sin\alpha\cos\alpha}{N_{u}^{2}%
}[\sigma_{ab}^{(2)}U^{(1)b}+\omega_{ab}^{(2)}U^{(1)b}+\frac{\theta_{(2)}}%
{3}(U_{a}^{(1)}+\digamma U_{a}^{(2)})]\nonumber
\end{align}

\subsubsection*{Acceleration with Shear-Free components}

When the components are shear-free, the expression for the composite
acceleration simplifies, with the vorticity as well as the shear terms
vanishing.\ To see this, consider the symmetric and antisymmetric combinations
of Eq.(\ref{B2}):%
\begin{align}
N_{u}(-\dot{U}_{a}U_{b}-\dot{U}_{b}U_{a}+2\sigma_{ab}+\frac{2\theta}{3}h_{ab})
&  =\cos\alpha\lbrack-\dot{U}_{a}^{(1)}U_{b}^{(1)}-\dot{U}_{b}^{(1)}%
U_{a}^{(1)}+2\sigma_{ab}^{(1)}+\frac{2\theta_{(1)}}{3}h_{ab}^{(1)}%
]\label{NU-1}\\
&  +B\sin\alpha\lbrack-\dot{U}_{a}^{(2)}U_{b}^{(2)}-\dot{U}_{b}^{(2)}%
U_{a}^{(2)}+2\sigma_{ab}^{(2)}+\frac{2\theta_{(2)}}{3}h_{ab}^{(2)}]\nonumber
\end{align}%
\begin{align}
N_{u}(-\dot{U}_{a}U_{b}+\dot{U}_{b}U_{a}+2\omega_{ab}) &  =\cos\alpha
\lbrack-\dot{U}_{a}^{(1)}U_{b}^{(1)}+\dot{U}_{b}^{(1)}U_{a}^{(1)}+2\omega
_{ab}^{(1)}]\label{NU-2}\\
&  +B\sin\alpha\lbrack-\dot{U}_{a}^{(2)}U_{b}^{(2)}+\dot{U}_{b}^{(2)}%
U_{a}^{(2)}+2\omega_{ab}^{(2)}]\nonumber
\end{align}
The acceleration can be obtained from either the symmetric or the
antisymmetric equation.\ Using both will provide the vorticity relation. First
using the symmetric combination, multiplying by $U^{b}$ we have%
\begin{align}
N_{u}\ \dot{U} &  =\cos\alpha(-\dot{U}_{a}^{(1)}U_{b}^{(1)}U^{b}-\dot{U}%
_{b}^{(1)}U^{b}U_{a}^{(1)}+2U^{b}\sigma_{ab}^{(1)}+\frac{2\theta_{(1)}}%
{3}h_{ab}^{(1)}U^{b})\\
&  +B\sin\alpha(-\dot{U}_{a}^{(2)}U_{b}^{(2)}U^{b}-\dot{U}_{b}^{(2)}U^{b}%
U_{a}^{(2)}+2U^{b}\sigma_{ab}^{(2)}+\frac{2\theta_{(2)}}{3}U^{b}h_{ab}%
^{(2)})\nonumber
\end{align}
Using the equation for $U^{b}$ in terms of $U^{b(1)}$ and $U^{b(2)}$, this can
be rewritten. \
\begin{align*}
\dot{U}_{a} &  =\dot{U}_{a}^{(1)}\cos^{2}\alpha+\dot{U}_{a}^{(2)}\sin
^{2}\alpha-\frac{B\sin\cos\alpha}{N_{u}^{2}}\ [\dot{U}_{b}^{(1)}U^{b(2)}%
U_{a}^{(1)}+\dot{U}_{b}^{(2)}U^{b(1)}U_{a}^{(2)}]\\
&  +\frac{2B\sin\alpha\cos\alpha}{N_{u}^{2}}(U^{b(1)}\sigma_{ab}%
^{(2)}+U^{b(2)}\sigma_{ab}^{(1)})\\
&  +\frac{B\sin\cos\alpha}{N_{u}^{2}}[\frac{2\theta_{(1)}}{3}(U_{a}%
^{(2)}+\digamma U_{a}^{(1)})+\frac{2\theta_{(2)}}{3}(U_{a}^{(1)}+\digamma
U_{a}^{(2)})]
\end{align*}
Using the same procedure with the anti-symmetric combination the acceleration
involving the component vorticity is found. \
\begin{align}
\dot{U}_{a} &  =\dot{U}_{a}^{(1)}\cos^{2}\alpha+\dot{U}_{a}^{(2)}\sin
^{2}\alpha\\
&  +\frac{B\sin\alpha\cos\alpha}{N_{u}^{2}}[\dot{U}_{b}^{(1)}U_{a}%
^{(1)}U^{b(2)}+\dot{U}_{b}^{(2)}U^{b(1)}U_{a}^{(2)}]\nonumber\\
&  +\frac{2B\cos\alpha\sin}{N_{u}^{2}}(\omega_{ab}^{(1)}U^{b(2)}%
+U^{b(1)}\omega_{ab}^{(2)})\nonumber
\end{align}
For the two results to be equal implies%
\begin{align}
&  U^{(1)b}\sigma_{ab}^{(2)}+U^{(2)b}\sigma_{ab}^{(1)}-U^{(2)b}\omega
_{ab}^{(1)}-U^{(1)b}\omega_{ab}^{(2)}\\
&  =\dot{U}_{b}^{(1)}U_{a}^{(1)}U^{(2)b}+\dot{U}_{b}^{(2)}U^{(1)b}U_{a}%
^{(2)}-\frac{\theta_{(1)}}{3}(U_{a}^{(2)}+\digamma U_{a}^{(1)})-\frac
{\theta_{(2)}}{3}(U_{a}^{(1)}+\digamma U_{a}^{(2)})\ \nonumber
\end{align}
Substituting for the acceleration velocity product from\ Eqs.(\ref{acc-vel-1}%
,\ref{acc-vel-2}) and combining, the component expansion contributions cancel
and we have
\begin{align*}
&  (\omega_{ab}^{(1)}-\sigma_{ab}^{(1)})U^{b(2)}+U^{b(1)}(\omega_{ab}%
^{(2)}-\sigma_{ab}^{(2)})\\
&  =\frac{U_{a}^{(1)}(\sigma_{ab}^{(1)}U^{a(2)}U^{b(2)}\digamma+\sigma
_{ab}^{(2)}U^{a(1)}U^{b(1)})+U_{a}^{(2)}(\sigma_{ab}^{(2)}U^{a(1)}%
U^{b(1)}\digamma+\sigma_{ab}^{(1)}U^{a(2)}U^{b(2)})}{(1-\digamma^{2})}%
\end{align*}
If the component shears are zero, this provides a relation between the
component vorticities \
\begin{equation}
\omega_{ab}^{(1)}U^{(2)b}=-U^{(1)b}\omega_{ab}^{(2)}%
\end{equation}
Substituting for the acceleration/velocity overlaps, the general composite
acceleration for shear-free components is
\begin{equation}
\dot{U}_{a}=\dot{U}_{a}^{(1)}\cos^{2}\alpha+\dot{U}_{a}^{(2)}\sin^{2}%
\alpha+\frac{B\sin\alpha\cos\alpha}{N_{u}^{2}}[\frac{\theta_{(1)}%
\digamma+\theta_{(2)}}{3}U_{a}^{(1)}+\frac{\theta_{(2)}\digamma+\theta_{(1)}%
}{3}U_{a}^{(2)}]
\end{equation}
The overlap of the composite acceleration onto the anisotropy vector for
shear-free components is%
\begin{equation}
\dot{U}_{a}\Upsilon^{a}=\frac{-\theta_{(1)}\sin\alpha+\theta_{(2)}B\cos\alpha
}{3BN_{\Upsilon}}%
\end{equation}
For expansion-free component fluids, this requires the acceleration and the
direction of anisotropy to be orthogonal. \ 

\subsubsection*{Fluid Shear}

The composite fluid shear is the most important of the composite parameters as
its form will set the component conditions for a shear-free composite.\ The
fluid shear follows from Eq.(\ref{B2}).\ The composite fluid tetrad is
$[U^{a},\Upsilon^{a},X^{a},Z^{a}],$where $\Upsilon^{a}$ is the transformation
anisotropy vector and $X^{a}$ and $Z^{a}$ are any other spatial vectors
orthogonal to $\Upsilon^{a}\ $and $U^{a}$.\ Any component velocity product
with $X^{a}$ or $Z^{a}$ is zero since the component velocities are rotated
into the pair $[U^{a},\Upsilon^{a}].$ Using the symmetric combination,
Eq.(B6), first multiply by $X^{a}$%
\begin{align*}
N_{u}(-\dot{U}_{a}X^{a}U_{b}+2X^{a}\sigma_{ab}+\frac{2\theta}{3}X_{b})  &
=\cos\alpha(-\dot{U}_{a}^{(1)}X^{a}U_{b}^{(1)}+2X^{a}\sigma_{ab}^{(1)}%
+\frac{2\theta_{(1)}}{3}X_{b})\\
&  +B\sin\alpha(-\dot{U}_{a}^{(2)}X^{a}U_{b}^{(2)}+2X^{a}\sigma_{ab}%
^{(2)}+\frac{2\theta_{(2)}}{3}X_{b})
\end{align*}
The composite expansion is related to the component expansions, Eq.(B3). Using
this we have%
\[
N_{u}(-\dot{U}_{a}X^{a}U_{b}+2X^{a}\sigma_{ab})=\cos\alpha(-\dot{U}_{a}%
^{(1)}X^{a}U_{b}^{(1)}+2X^{a}\sigma_{ab}^{(1)})+B\sin\alpha(-\dot{U}_{a}%
^{(2)}X^{a}U_{b}^{(2)}+2X^{a}\sigma_{ab}^{(2)})
\]
Multiplying by the unit vectors $X^{b}$ and $Z^{b}$%
\begin{align}
N_{u}X^{a}X^{b}\sigma_{ab}  &  =X^{a}X^{b}(\cos\alpha\sigma_{ab}^{(1)}%
+B\sin\alpha\sigma_{ab}^{(2)})\\
N_{u}X^{a}Z^{b}\sigma_{ab}  &  =X^{a}Z^{b}(\cos\alpha\sigma_{ab}^{(1)}%
+B\sin\alpha\sigma_{ab}^{(2)})
\end{align}
The shear component in the $Z^{a}Z^{a}$ direction follows directly by analogy%
\begin{equation}
N_{u}Z^{a}Z^{b}\sigma_{ab}=Z^{a}Z^{b}(\cos\alpha\sigma_{ab}^{(1)}+B\sin
\alpha\sigma_{ab}^{(2)})
\end{equation}
For shear-free components, these composite shear components will be
zero.\ First repeating with a $\Upsilon^{b}$ contraction
\begin{align*}
N_{u}(-\dot{U}_{b}\Upsilon^{b}U_{a}+2\Upsilon^{b}\sigma_{ab}+\frac{2\theta}%
{3}\Upsilon_{a})  &  =\cos\alpha\lbrack-\dot{U}_{a}^{(1)}\Upsilon^{b}%
U_{b}^{(1)}-\dot{U}_{b}^{(1)}\Upsilon^{b}U_{a}^{(1)}+2\Upsilon^{b}\sigma
_{ab}^{(1)}+\frac{2\theta_{(1)}}{3}\Upsilon^{b}h_{ab}^{(1)}]\\
&  +B\sin\alpha\lbrack-\dot{U}_{a}^{(2)}\Upsilon^{b}U_{b}^{(2)}-\dot{U}%
_{b}^{(2)}\Upsilon^{b}U_{a}^{(2)}+2\Upsilon^{b}\sigma_{ab}^{(2)}+\frac
{2\theta_{(2)}}{3}\Upsilon^{b}h_{ab}^{(2)}]
\end{align*}
and then $X^{b}$ and $Z^{b}$
\begin{subequations}
\label{many-shear}%
\begin{align}
X^{b}\Upsilon^{a}\sigma_{ab}  &  =\frac{B\sin\alpha\cos\alpha N_{\Upsilon}%
}{2N_{u}}X^{b}(\dot{U}_{b}^{(1)}-\dot{U}_{b}^{(2)})+\frac{X^{b}\Upsilon^{a}%
}{N_{u}}(\cos\alpha\sigma_{ab}^{(1)}+B\sin\alpha\sigma_{ab}^{(2)})\\
Z^{b}\Upsilon^{a}\sigma_{ab}  &  =\frac{B\sin\alpha\cos\alpha N_{\Upsilon}%
}{2N_{u}}Z^{b}(\dot{U}_{b}^{(1)}-\dot{U}_{b}^{(2)})+\frac{Z^{b}\Upsilon^{a}%
}{N_{u}}(\cos\alpha\sigma_{ab}^{(1)}+B\sin\alpha\sigma_{ab}^{(2)})
\end{align}
The last shear component $\Upsilon^{a}\Upsilon^{b}\sigma_{ab}$ can be
calculated and it is where the shear-free conditions enter\
\end{subequations}
\begin{align}
N_{u}(2\Upsilon^{a}\Upsilon^{b}\sigma_{ab}+\frac{2\theta}{3})  &  =\cos
\alpha\lbrack(-\dot{U}_{a}^{(1)}\Upsilon^{a}\Upsilon^{b}U_{b}^{(1)}-\dot
{U}_{b}^{(1)}\Upsilon^{a}\Upsilon^{b}U_{a}^{(1)}\label{B15}\\
&  +2\Upsilon^{a}\Upsilon^{b}\sigma_{ab}^{(1)}+\frac{2\theta_{(1)}}%
{3}(1+\Upsilon^{b}U_{b}^{(1)}\Upsilon^{a}U_{a}^{(1)})]\nonumber\\
&  +B\sin\alpha\lbrack(-\dot{U}_{a}^{(2)}\Upsilon^{a}\Upsilon^{b}U_{b}%
^{(2)}-\Upsilon^{b}\dot{U}_{b\ }^{(2)}U_{a}^{(2)}\Upsilon^{a}+2\Upsilon
^{a}\Upsilon^{b}\sigma_{ab}^{(2)}\nonumber\\
&  +\frac{2\theta_{(2)}}{3}(1+\Upsilon^{b}U_{b}^{(2)}\Upsilon^{a}U_{a}%
^{(2)})]\nonumber
\end{align}
Using Eq.(\ref{B3}), the expansion terms on the left side of the equation
cancel and we have%
\begin{align}
N_{u}\ \Upsilon^{a}\Upsilon^{b}\sigma_{ab}  &  =N_{\Upsilon}B\sin\alpha
\cos\alpha\ [\Upsilon^{a}(\dot{U}_{a}^{(1)}-\dot{U}_{a}^{(2)})+N_{\Upsilon
}\frac{\theta_{(2)}\cos\alpha+\theta_{(1)}B\sin\alpha}{3}\label{N-Gam-Gam}\\
&  +\Upsilon^{a}\Upsilon^{b}(\cos\alpha\ \sigma_{ab}^{(1)}+B\sin\alpha\text{
}\sigma_{ab}^{(2)})]\nonumber
\end{align}
The shear has zero trace and since $\sigma_{ab}X^{a}X^{b}$ and $\sigma
_{ab}Z^{a}Z^{a}=0,$ for shear-free components this requires $\Upsilon
^{a}\Upsilon^{b}\sigma_{ab}=0.$ A composite shear-free condition is
\begin{equation}
\Upsilon^{a}(\dot{U}_{a}^{(2)}-\dot{U}_{a}^{(1)})=N_{\Upsilon}\frac{\theta
_{2}\cos\alpha+\theta_{1}B\sin\alpha}{3}\
\end{equation}

\subsubsection*{Vorticity}

The vorticity follows from the antisymmetrized Eq.(\ref{NU-2}).\ The
$\Upsilon^{a}X^{b}$ contraction is%
\begin{equation}
\Upsilon^{a}X^{b}\omega_{ab}=[\frac{B\sin\alpha\cos\alpha N_{\Upsilon}}%
{2N_{u}}]X^{b}(\dot{U}_{b}^{(2)}-\dot{U}_{b}^{(1)})+\frac{\Upsilon^{a}X^{b}%
}{N_{u}}(\cos\alpha\omega_{ab}^{(1)}+B\sin\alpha\omega_{ab}^{(2)})
\end{equation}
and similarly%
\begin{align*}
\Upsilon^{a}Z^{b}\omega_{ab}  &  =[\frac{B\sin\alpha\cos\alpha N_{\Upsilon}%
}{2N_{u}\ }]Z^{b}(\dot{U}_{b}^{(2)}-\dot{U}_{b}^{(1)})+\frac{\Upsilon^{a}%
Z^{b}}{N_{u}}(\cos\alpha\omega_{ab}^{(1)}\ +B\sin\alpha\omega_{ab}^{(2)})\\
Z^{a}X^{b}\omega_{ab}  &  =\frac{Z^{a}X^{b}}{N_{u}}(\cos\alpha\omega
_{ab}^{(1)}+B\sin\alpha\omega_{ab}^{(2)})
\end{align*}

\section{Fluid Acceleration and Expansion for shear-free components}

The component fluids that will combine to create a shear-free composite have
strong restrictions on their accelerations and expansions. The shear of the
composite fluid is the most important of the composite parameters as its form
will set the component conditions.\ From Eqs.(\ref{many-shear},\ref{N-Gam-Gam}%
), for shear-free components, the composite shear tensor components are
\begin{subequations}
\label{shear-comp}%
\begin{align}
X^{a}X^{b}\sigma_{ab}  &  =Z^{a}Z^{b}\sigma_{ab}=X^{a}Z^{b}\sigma_{ab}=0\\
X^{a}\Upsilon^{b}\sigma_{ab}  &  =[\frac{B\sin\alpha\cos\alpha N_{\Upsilon}%
}{2N_{u}}]X^{a}(\dot{U}_{a}^{(1)}-\dot{U}_{a}^{(2)})\\
Z^{a}\Upsilon^{b}\sigma_{ab}  &  =[\frac{B\sin\alpha\cos\alpha N_{\Upsilon}%
}{2N_{u}}]Z^{a}(\dot{U}_{a}^{(1)}-\dot{U}_{a}^{(2)})\\
\Upsilon^{a}\Upsilon^{b}\sigma_{ab}  &  =\frac{N_{\Upsilon}}{N_{u}}B\sin
\alpha\cos\alpha\left[  \Upsilon^{a}(\dot{U}_{a}^{(1)}-\dot{U}_{a}%
^{(2)})+N_{\Upsilon}(\frac{\theta_{(2)}\cos\alpha+\theta_{(1)}B\sin\alpha}%
{3})\right]
\end{align}
An inspection of the composite shear establishes two conditions for the
composite fluid to be shear-free.%
\end{subequations}
\begin{align}
X^{a}(\dot{U}_{a}^{(1)}-\dot{U}_{a}^{(2)})  &  =Z^{a}(\dot{U}_{a}^{(1)}%
-\dot{U}_{a}^{(2)})=0\\
\Upsilon^{a}(\dot{U}_{a}^{(2)}-\dot{U}_{a}^{(1)})  &  =N_{\Upsilon}%
(\frac{\theta_{(2)}\cos\alpha+\theta_{(1)}B\sin\alpha}{3})
\end{align}
For aligned fluids, $U_{a}^{(1)}=U_{a}^{(2)},$ these are identities with equal
accelerations and $\alpha=\alpha_{0}.$\ Shear-free aligned fluids will produce
a shear-free composite but it is isotropic and has no heat flow.\ For
unaligned fluids, two possible conditions are equal or unequal component
accelerations.\ The fluid overlap, $\digamma,$ is assumed constant and
establishes a relation between the component accelerations.\ From Appendix A
we have \
\begin{align}
\dot{U}_{a}^{(1)}U^{(2)a}  &  =(\digamma\frac{\theta_{(1)}}{3}+\frac
{\theta_{(2)}}{3})\\
\dot{U}_{a}^{(2)}U^{(1)a}  &  =(\frac{\theta_{(1)}}{3}+\digamma\frac
{\theta_{(2)}}{3})
\end{align}
Equal component acceleration requires $\theta_{1}\digamma=-\theta_{2}$ and
$\theta_{2}\digamma=-\theta_{1}.$ This implies the component fluids are
aligned, $\digamma^{2}=1$.\ Two unaligned fluids with equal accelerations will
not produce a shear-free composite.\ The fluid accelerations should have
non-canceling components along the direction of anisotropy.

\end{document}